\documentclass{article}
\usepackage[a4paper, total={7in, 10in}]{geometry}
\usepackage[table]{xcolor}

\usepackage{colortbl}
\usepackage[hyphens]{url}
\usepackage{subcaption}
\usepackage{graphicx}
\usepackage{amssymb}
\usepackage{booktabs}
\usepackage{multirow}
\usepackage{color}
\usepackage{rotating}
\usepackage{nicematrix}
\providecommand{\keywords}[1]{\textbf{\textit{Index terms---}} #1}
\graphicspath{{images/}}
\begin{document}
\title{The Cynicism of Modern Cybercrime: Automating the Analysis of Surface Web Marketplaces}
\author{Nikolaos Lykousas$^{1}$, Vasilios Koutsokostas$^{1}$, Fran Casino$^{1,2}$ and Constantinos Patsakis$^{1,2}$\\
        \small $^{1}$Department of Informatics, University of Piraeus,\\\small 80 Karaoli \& Dimitriou str., 18534 Piraeus, Greece\\
        \small $^{2}$Information Management Systems Institute,\\\small Athena Research Center, Artemidos 6, Marousi 15125, Greece\\
}
\date{}
\maketitle

\begin{abstract}
Cybercrime is continuously growing in numbers and becoming more sophisticated. Currently, there are various monetisation and money laundering methods, creating a huge, underground economy worldwide. A clear indicator of these activities is online marketplaces which allow cybercriminals to trade their stolen assets and services. While traditionally these marketplaces are available through the dark web, several of them have emerged in the surface web.

In this work, we perform a longitudinal analysis of a surface web marketplace. The information was collected through targeted web scrapping that allowed us to identify hundreds of merchants' profiles for the most widely used surface web marketplaces. In this regard, we discuss the products traded in these markets, their prices, their availability, and the exchange currency. This analysis is performed in an automated way through a machine learning-based pipeline, allowing us to quickly and accurately extract the needed information. The outcomes of our analysis evince that illegal practices are leveraged in surface marketplaces and that there are not effective mechanisms towards their takedown at the time of writing.
\end{abstract}

\section{Introduction}
With the continuous digitisation of procedures, services, and products, crime has shifted towards the same direction. As a result, almost every aspect of a modern crime is facilitated by digital means, and consequently, almost every criminal investigation involves some sort of digital evidence. The above are the primary reasons why cybercrime has evolved into a multi-billion underground economy. Its economic impact is devastating \cite{thomas2020cybercrime}, with FBI estimating the losses to \$3.5 billion only within USA \cite{crime_report}. In fact, this continuous rise is so threatening that it has become the second most-concerning risk for global commerce over the next decade, according to the World Economic Forum \cite{wef}. Notably, the recent COVID-19 pandemic resulted in a huge spike in cybercrime activities\footnote{\url{https://www.europol.europa.eu/newsroom/news/covid-19-sparks-upward-trend-in-cybercrime} \url{https://www.interpol.int/en/News-and-Events/News/2020/INTERPOL-report-shows-alarming-rate-of-cyberattacks-during-COVID-19}} making the problem even more thorny.
It is indisputable that the introduction of cryptocurrencies has significantly facilitated illegal money flows. The transactions of many cryptocurrencies offer high inherent privacy guarantees; e.g. Monero, ZCash, or they are difficult to be traced due to the wallet anonymity; e.g. Bitcoins. The rise of the dark web has also boosted the underground economy as it provides perpetrators with more advanced privacy guarantees.

While dark web markets are still the key stakeholders when it comes to illegal trading, several surface web marketplaces have recently been repeatedly reported for trading leaked data\footnote{\url{https://www.forbes.com/sites/zakdoffman/2020/04/20/facebook-users-beware-hackers-just-sold-267-million-of-your-profiles-for-540/} \url{https://www.zdnet.com/article/hacker-selling-data-of-538-million-weibo-users/}}. A very interesting characteristic, in this case, is that despite the trading of illicit products, these marketplaces have a very open form, e.g. they do not require any registration to access them, and the ``loot'' that is traded is advertised openly across the web.


\subsection{Motivation and Contribution}
Delinquent behaviour is unarguably a characteristic that can be observed in every human society. The extent of this behaviour, in terms of how many people exhibit it, and the harm that is caused by it define the ethics of the society and its limits.

The ethics on the Internet follow different rules \cite{spinello2010cyberethics}. Moreover, the dark web is considered the default place on the Internet in which such behaviours and actions flourish. Nonetheless, they are promoted in closed circles so that there is some ``control'' on who can access this information and to retain the anonymity of the perpetrators. However, should this information be openly disseminated in public channels, it implies that the promoted behaviour is widely practised and is considered a norm by some groups.

In the past few years, there has been a significant increase in reported data leaks, online extortion schemes and credential trading. One of our initial research questions was whether such actions are so widely performed that they can be observed on the surface web. In this regard, we wanted to check whether the perpetrators were using platforms of the surface web to advertise their ``loot'' and the existence of marketplaces in the surface web.

Currently, there are several such marketplaces operating with similar functionality; however, this work is mainly focused on \texttt{Shoppy} (\url{https://shoppy.gg/}) which appears to have the most users and products at the time of writing. Nonetheless, similar illicit trends have been found in the rest of the surface web marketplaces. The goal of this work is to provide an overview of what is actually being sold in such a marketplace, and leverage methods (e.g. machine learning) to automatically determine which are the illegal products and the main organisations affected. The main limitation in the automation of such a task is the lack of text. These sellers do not need to add a lot of text about what they trade in these marketplaces, and in many occasions there are typos, abbreviations, and slang, posing even more issues in the analysis of the derived text. Further to the analysis of the traded products, we discuss the modus operandi of the sellers and some insight regarding the pricing of ``big leaked data''.

The rest of this work is structured as follows. First, we provide an overview of the related work and a brief discussion of these marketplaces. Then, we detail our data collection methodology. In Section \ref{sec:explore}, we analyse the collected data to extract actual knowledge out of the short descriptions of the shops in an automated way. Finally, the article concludes, summarising our contributions and highlighting some ideas for future work.

\section{Related work}

In recent years, there have been multiple incidents of massive data breaches affecting a broad spectrum of online services and service providers, including retailers, payment processors, and government entities \cite{decary2016criminals}. Malicious actors gain internal access to sensitive data sources, and then acquire millions of credit and debit card details, user credentials, as well as sensitive data which can be used to identify individuals uniquely. The sheer quantity of data that can be acquired has given rise to a burgeoning market for actors who sell the information that they obtain, through, e.g. hacking and other forms of data theft, to other users. Participants in these illegally acquired data markets leverage various communication and networking methods, enabling them to freely form communities and interaction mechanisms. The most prevalent forms of such marketplaces, as identified in the literature, are Internet forums and Internet-Relay-Chat (IRC) channels \cite{peretti2008data,benjamin2015exploring}.
In particular, forums have been shown to comprise the principal medium for cybercriminals to network, form communities, and operate online stolen data markets, despite numerous successful infiltrations by law enforcement agencies \cite{yip2013forums}.
To a large extent, these marketplaces reside in the dark web, commonly behind Tor \cite{dingledine2004tor}, and are referred to as ``Darknet markets''. Darknet markets are popular among criminals since they enable them to anonymously trade illegal goods and services, extending well beyond stolen data. The latter was discussed by Thomas et al. \cite{thomas2015framing} by pointing out the complex value chain of the underground market economy at scale. These marketplaces comprise the essential pillars of this global-scale cybercrime economy and thus have become the key information source for investigating the cybercriminal ecosystem. An extensive body of literature has explored the darknet marketplaces \cite{wehinger2011dark}, the involved stakeholders and their communication patterns \cite{hutchings2015crime}, and their modus operandi \cite{wang2018you}.

Of particular interest to researchers, are the marketplaces dedicated to the sale of stolen personal and financial information, known as ``carding forums'', where cybercriminals sell the artefacts of large scale data breaches, often containing stolen financial information \cite{peretti2008data,li2016identifying,haslebacher2017all}. Subsequently, the compromised credit and debit card information enables malicious actors to commit crimes such as identity theft, financial fraud, and most importantly, online money laundering \cite{decary2016criminals,mikhaylov2016cards}.
Moreover, due to their illicit and underground nature, carding forums and marketplaces are characterised by unique trading dynamics between vendors and sellers, since the quality of merchandise and the identities of traders are unknown to potential buyers. In this regard, a number of works focus on untangling the mechanics of transactions in this particular kind of underground marketplaces \cite{yip2013trust,haslebacher2015understanding, holt2016examining,sundaresan2016profiling}.

Identifying key players is essential when investigating emergent threats and developing efficient disruption strategies \cite{park2018hackers}, in particular, considering the fact that members of such communities are characterised by cross-forum posting activity, which can be used to identify user roles based on the type of posts and their frequencies. The indicators of trustworthiness and reputation of a seller play a pivotal role for the sale of illicit services and stolen data through underground hacking forums and markets, as users are more likely to conduct business with sellers who hold reputable standing. As such, it has been shown that the status of reputation can be used to identify prominent players in illicit online marketplaces \cite{yip2013trust}.

Apart from the complexity of its dynamics, the ecosystem of illegal online markets is characterised by an equally wide range of offerings, services and products relevant for a variety of illicit topics, such as underground drug economies, data breaches, and cyber warfare.
For instance, the work proposed in \cite{1394049} focuses on malicious assets traded in hacker forums, such as hacking tools, rootkits and exploits, from which cyber threat intelligence can be distilled, and analyses the extracted data for predicting and mitigating cyber attacks. In \cite{7745465}, the authors provide both quantitative and qualitative categorisation of offerings in 17 different marketplaces. Their findings indicate the existence of both highly specialised products with respect to particular vendors and markets, as well as the cross-listing of products on multiple sites and nearly identical products for sale by multiple vendors. Nonetheless, the most prevalent categories are related to stolen credentials and information, extending beyond financial accounts. In this direction, Madarie et al. \cite{madarie2019stolen} examined how, a diverse set of outlets such as stolen credentials, are disseminated by malicious actors as ``account dumps’’. Their analysis revealed that the illicit dissemination of stolen account credentials covers a broad spectrum of online services, and highlighted Pastebin\footnote{\url{https://pastebin.com/}} as one of the main sites used to spread the information. Pastebin is a paste website intended for sharing plain text snippets, which is radically different compared to the typical marketplaces and forums, due to the complete lack of structure. Interestingly enough, while stolen combinations of usernames and passwords for various online services were posted, data thieves used advertisements (embedded within the dumps) for establishing communication with potential customers seeking access to financial or more sensitive data.


\section{Methodology}

After thorough literature research, we observed a literature gap focusing on surface web marketplaces and an analysis of some of their activities, including automated methodologies for the extraction and processing of information to discover potentially malicious behaviour.
First, we explored two well-known hacking-related forums, namely \texttt{blackhatworld}\footnote{https://www.blackhatworld.com/} and \texttt{cracked}\footnote{https://cracked.to/} looking for indications of the emergence of marketplaces supporting illegal activities, similar to the deep web forums/marketplaces. In this regard, we observed that \texttt{Shoppy}\footnote{https://shoppy.gg/} was widely used in such platforms to monetise some of the reported activities. In fact, such activities were advertised. Therefore, we established a methodology to analyse what types of activities and products were being sold in \texttt{Shoppy}. \texttt{Shoppy} is a shop hosting service that provides the opportunity to individual vendors to sell their products, allows payments in different forms, and a set of APIs to, e.g. advertise one's products in forums etc. 
A crucial difference between \texttt{Shoppy} and the underground marketplaces studied in the literature is that the former does not offer a centralised listing of the sold vendors and products. Each vendor obtains a unique URL where they can host their shop, without providing any means for a user to look for similar shops of products offered by different vendors, a common feature in e-commerce platforms \cite{schafer2001commerce}.
The decentralised architecture of \texttt{Shoppy} hinders the extraction of knowledge, and thus, our proposed data collection methodology specifically aims to discover shops associated with illicit offerings and services, given the context established by focusing on hacking-related forums.

\begin{figure}
 \centering
 \includegraphics[trim={0 3.8cm 0 0},clip,width=\textwidth]{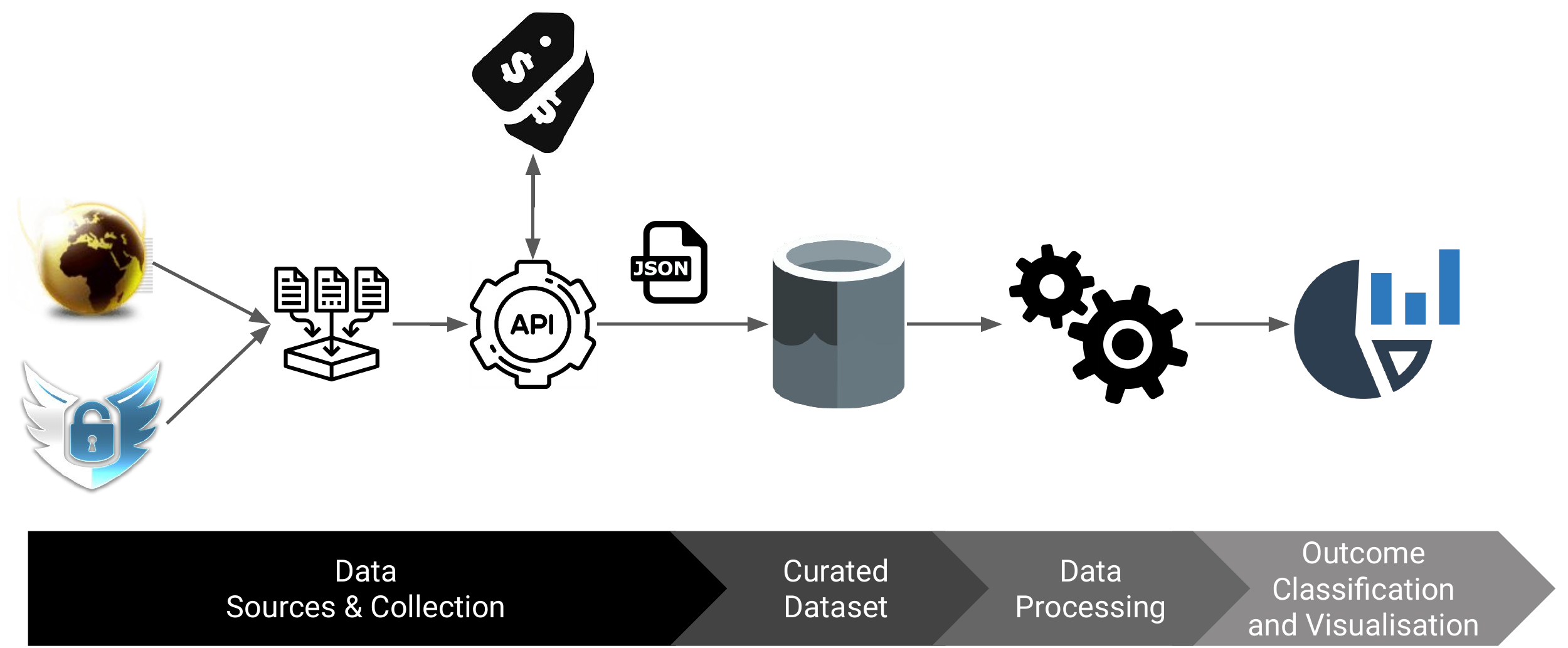}
 \caption{Workflow of the steps followed in our research.}
 \label{fig:methodology}
\end{figure}

The methodology that we adopted to address this challenge consists of several steps, as depicted in Figure \ref{fig:methodology}. First, we crawled the \texttt{blackhatworld} and \texttt{cracked} forums, collecting usernames, as well as references to \texttt{Shoppy} accounts in post signatures. Given the size of these two communities, we specifically focused our crawling only to the ``Marketplace'' forums. To his end, we adopted the architecture of the Structure-driven Incremental Forum crawler (SInFo) \cite{pavkovic2019sinfo}, which enabled us to crawl data from the aforementioned forums. Nevertheless, we did not leverage user accounts that could potentially allow us to access even more content, restricted to authenticated users \cite{tavabi2019characterizing}. Next, we examined the extent to which the collected usernames and \texttt{Shoppy} account data could be correlated with existing shops in the \texttt{Shoppy} ecosystem.
The data collection process lasted from March to April of 2020. We collected a total of 68,045 usernames, and \texttt{Shoppy} links from forum post signatures, 2,906 of which were linked to existing \texttt{Shoppy} shops at the time of crawling. The results are summarised in Table \ref{tab:crawling}. Notably, a large fraction of the links to \texttt{Shoppy} accounts found in post signatures, that did not resolve to existing shops, indicating that accounts in \texttt{Shoppy} may be banned, deleted, or renamed.

\begin{table}[th]
 \centering
\begin{tabular}{lrr}
\toprule
\textbf{Source}& \textbf{\#} & \textbf{Valid}\\
\midrule
blackhatworld - usernames & 24,658 & 827\\
blackhatworld - signatures & 660 & 359 \\
cracked.to - usernames & 41,890 & 1,230\\
cracked.to - signatures & 837 & 490 \\
\midrule
\textbf{Total (unique)} & 64,726 & 2,906 \\
\bottomrule
\end{tabular}
\caption{Collected usernames and \texttt{Shoppy} links.}
\label{tab:crawling}
\end{table}

With the collected data, we used the open \texttt{Shoppy} API to retrieve all the information associated with these shops, including products, prices, and their corresponding metadata to create a curated dataset.

\section{Data Exploration}
\label{sec:explore}
In the following sections, we explore the \texttt{Shoppy} data in different steps. First, we provide a quantitative review of the collected dataset. Next, we detail our topic modelling approach and, finally, we leverage an exploratory analysis of a subset of the surface data.

\subsection{Shoppy in Numbers}

In this section, we provide a quantitative analysis of the collected \texttt{Shoppy} shops and advertised products, as well as highlight the particular behaviours of vendors. In total, our dataset contains 64,726 products advertised by 2,906 vendors.
\texttt{Shoppy} provides vendors with the ability to categorise their products as accounts, services or files. The distribution of product categories in our dataset is provided in Table \ref{tab:prods}. ``Account'' is the default category, which evidently dominates the other two by a large margin.

\begin{table}[th]
 \centering
\begin{tabular}{lr}
\toprule
\textbf{Type}& \textbf{Count}\\
\midrule
Account & 52,850 \\
Service & 8,708 \\
File & 3,168 \\
\midrule
\textbf{Total} & 64,726 \\
\bottomrule
\end{tabular}
\caption{Shoppy products per category}
\label{tab:prods}
\end{table}

\begin{figure*}[!ht]
 \centering
 \begin{subfigure}[t]{0.49\textwidth}
 	\centering
 \includegraphics[width=\textwidth]{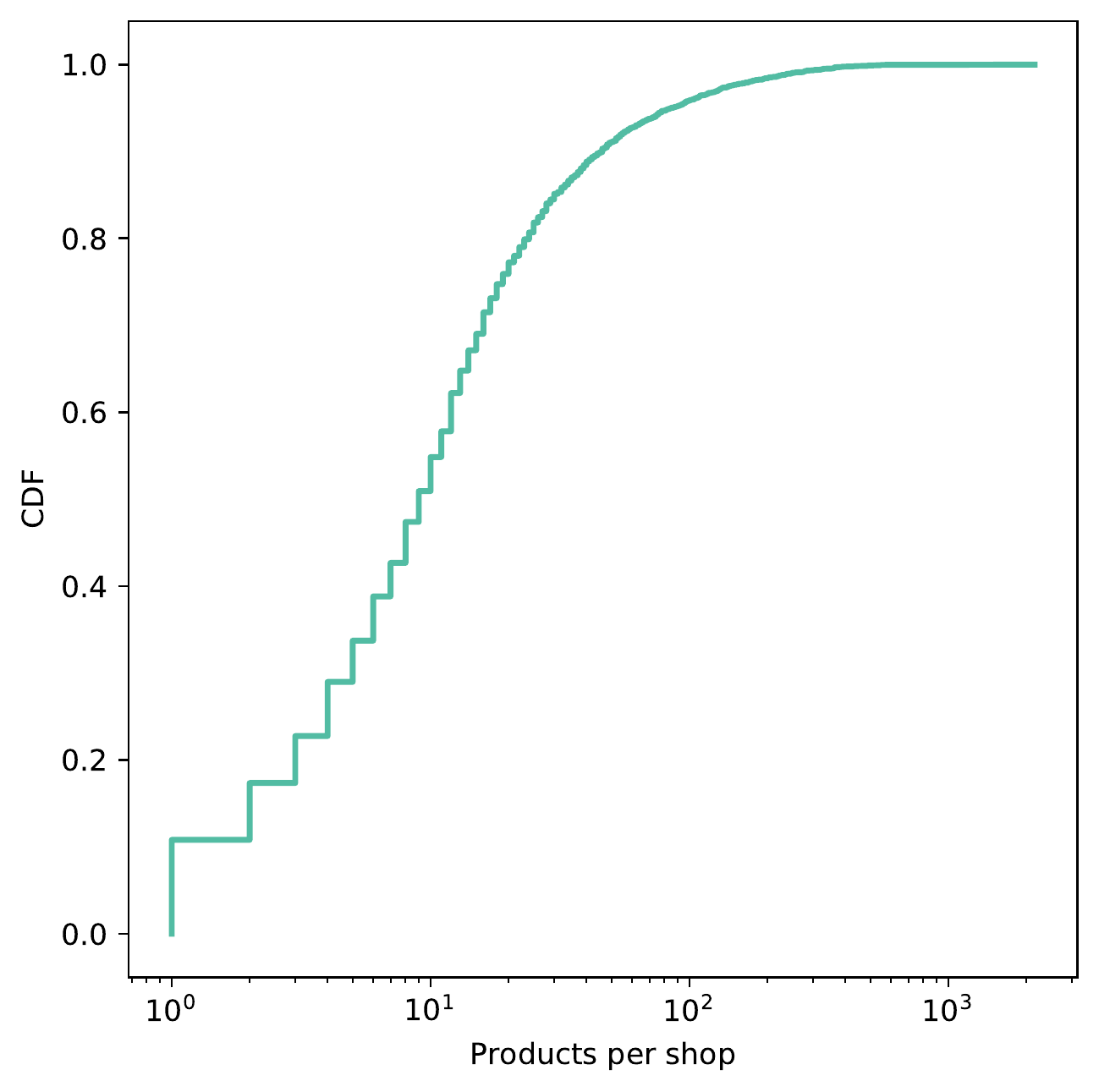}
 \caption{Products per shop}
 \label{fig:products_cdf}
 \end{subfigure}
 \begin{subfigure}[t]{0.49\textwidth}
 	\centering
 \includegraphics[width=\textwidth]{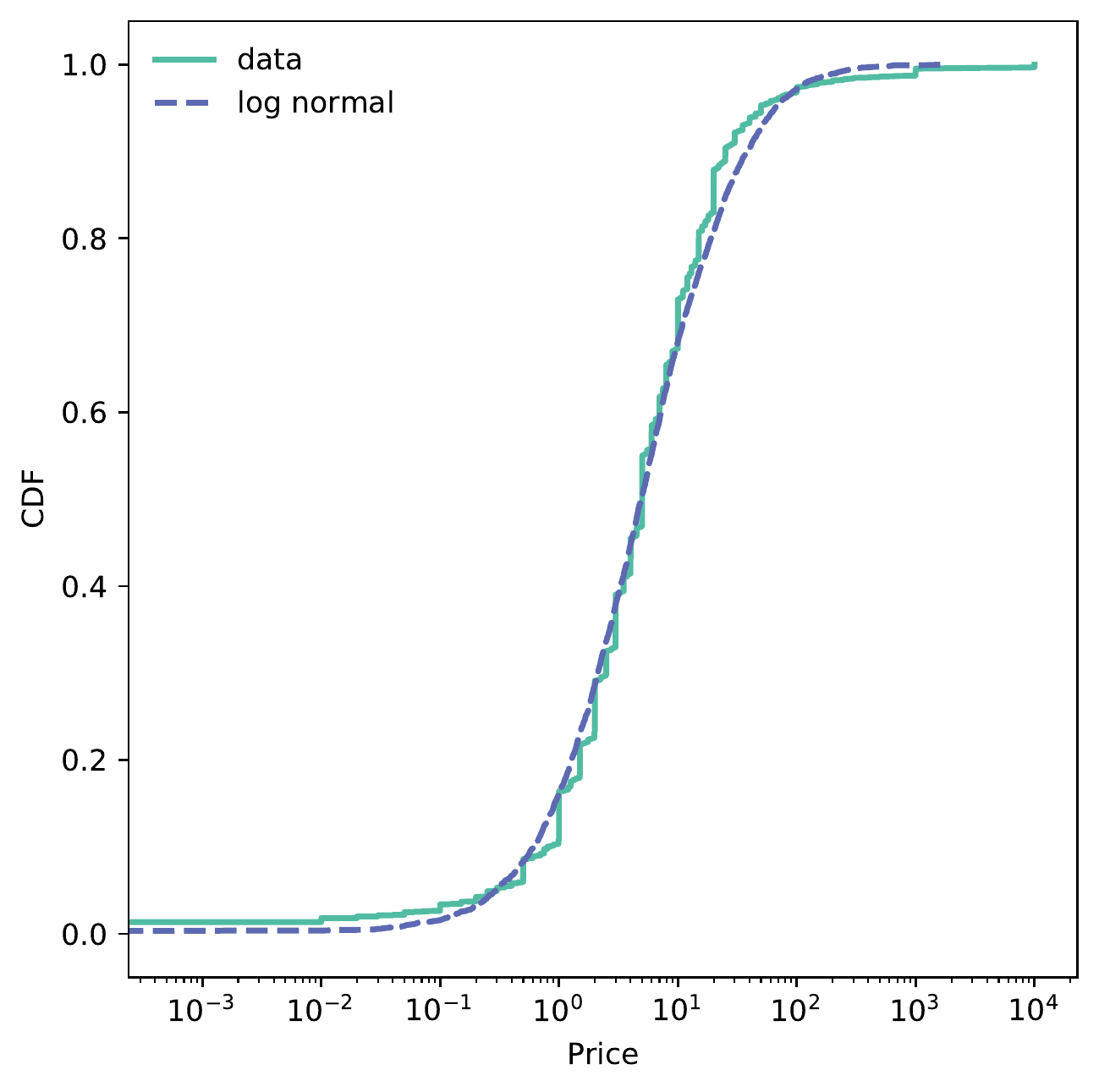}
 \caption{Price in USD}
 \label{fig:price_cdf}
 \end{subfigure}
 \caption{CDFs}
 \label{fig:cdfs}
\end{figure*}

Fig~\ref{fig:cdfs} describes the cumulative distribution functions (CDFs) of the number of items per shop (Fig~\ref{fig:products_cdf}) and the product prices, in USD (Fig~\ref{fig:price_cdf}).
We can observe that, while around 40\% of the shops have less than ten items listed, there exist shops with thousands of items. As seen in Fig \ref{fig:price_cdf}, the price distribution of products is remarkably well described by a lognormal distribution ($\mu=1.6, \sigma=1.58$), highlighting that the prices of approximately 62\% of the products fall within a small range comprised between 1 to 10\$. Moreover, the median price of \texttt{Shoppy} products is 5 USD, and, as observed in our dataset, the prices can reach up to 10,000 USD.

\begin{figure}[th]
 \centering
 \includegraphics[width=.8\textwidth]{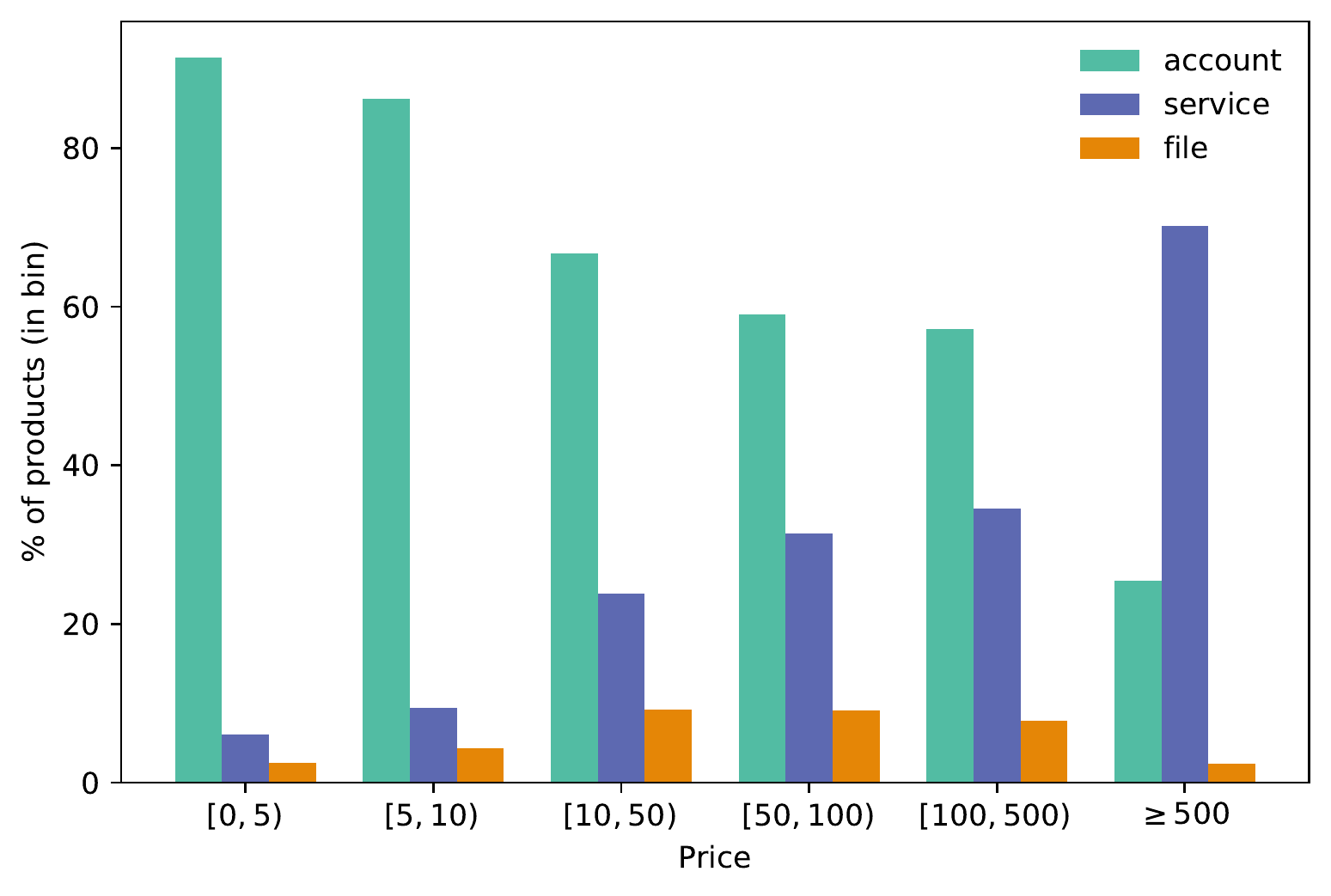}
 \caption{Fractional price bins of Shoppy products.}
 \label{fig:frac_bins}
\end{figure}

To get deeper insight on how different types of products are priced, focusing on possible outliers priced well above the median of 5 USD, we bin the products based on 
their price and we calculate the fractions of each product type in each bin in Figure~\ref{fig:frac_bins}.
We can observe that while the lowest price bin is dominated by accounts, the fractions of services per bin follow a consistently increasing trend as the prices increase. In contrast,
the relative representation of accounts is inversely proportional to the price, ultimately making services the predominant product category (approx. 70\% of total) for the last bin reflecting the highest-priced offerings ($\geq 500 \$$).
The fractions of the file type products, which as previously shown comprise only a small fraction of the total offerings, are generally sustained, accounting for less than 10\% of the products in each bin. It is worth to note that our initial observation related with the use of default categories is reflected in Figure \ref{fig:frac_bins}, which shows that, for instance, account products are well represented within all the range of possible prices. The latter behaviour seems quite unrealistic in a real and competitive market scenario and is further supported by the experiments performed in the next sections.


To investigate the high priced services dominating the upper price bracket, we manually examined them and provided some illustrative examples in Table \ref{tab:false_prods}.
Evidently, these items are false products and rather contain information such as merchants' terms of service, notes regarding provided shop feedback, support information, and links to Discord servers and Telegram channels maintained by the merchants. This behaviour has been highlighted in recent literature by arguing that the unregulated and anonymous nature of platforms such as Telegram and Discord, makes them the perfect habitats for scammers and cybercriminals \cite{nizzoli2020charting,turk2020tight}.

\begin{table}
\centering
	\begin{tabular}{p{.9\columnwidth}}
 \toprule
 	 {\bf Product Title} \\
 \midrule
 Terms of Service. READ BEFORE BUYING \\
 Terms of Service \& General Information \\
 Discord - DONT BUY \\
 Come Join! | Premium Town | Our Newest Discord Server \\
 Discord, Telegram, Skype Support Group (Special Discount Codes) \\
 Terms Of Service / Warranty Information \\
 Attention Contact me for support on telegram \\
 Discord \& Telegram server Links + Website Link [Click Me] \\
 Discord Server | Join For Support IMPORTANT - NEW DISCORD 13.04.2020 banned again \\
 \bottomrule
 \end{tabular}
\caption{Some illustrative false ``services'', priced $\geq 500 \$$. }
\label{tab:false_prods}
\end{table}


\subsection{Topic Modelling}

In this section, we analyse the \texttt{Shoppy} stores and elaborate a topic-based characterisation of the offered products by analysing their titles.
To this end, we consider a statistical model, namely ``topic model'', which is a method well suited to the study of high-level relationships between text documents.
Specifically, we leverage Latent Dirichlet Allocation (LDA), a generative probabilistic model proposed in \cite{blei2003latent}. It comprises an endogenous NLP technique, which as highlighted in \cite{cambria2014jumping} ``\textit{involves the use of machine-learning techniques to perform semantic analysis of a corpus by building structures that approximate concepts from a large set of documents}'' without relying on any external knowledge base. LDA, as the name implies, is a latent variable model in which each item in a collection (e.g., each text document in a corpus) is modelled as a finite mixture over an underlying set of topics. Each of these topics is characterised by a distribution over item properties (e.g. words). LDA assumes that these properties are exchangeable (i.e. ordering of words is ignored, as in many other ``bag of words'' approaches in text modelling), and that the properties of each document are observable (e.g. the words in each document are known). The word distribution for each topic and the topic distribution for each document are unobserved; they are learned from the data.

Since LDA is an unsupervised topic modelling method, there is no direct measure to identify the optimal number of topics to include in a model. In this sense, LDA assigns documents to different clusters of topics with certain probabilities (i.e. the number of clusters is defined with an integer number $k$ provided by the user), where these probabilities depend on the occurrence of words which are assumed to co-occur in documents belonging to the same topic (Dirichlet prior assumption). This exemplifies the main idea behind all unsupervised topic models, that language is organised by latent dimensions that actors may not even be aware of \cite{mcfarland2013differentiating}. 
Researchers have recommended various approaches to establish the optimal $k$ (e.g. \cite{cao2009density,arun2010finding,deveaud2014accurate,roder2015exploring,zhao2015heuristic}).
These approaches provide a good range of possible $k$ values that are mathematically plausible. However, according to \cite{dimaggio2013exploiting}, when topic modelling is used to identify themes and assist in interpretation (like in the present study), rather than to predict a knowable state or quantity, there is no statistical test for the optimal number of topics or the quality of a solution. A simple way to evaluate topic models is to look at the qualities of each topic and discern whether they are reasonable \cite{mcfarland2013differentiating}. 
To the best of our knowledge, the topic coherence measure with the largest correlation to human interpretability is the $C_v$ score defined in \cite{roder2015exploring}, which we also adopt in this study to establish the optimal number of topics.

\begin{figure}[th]
 \centering
 \includegraphics[width=.5\textwidth]{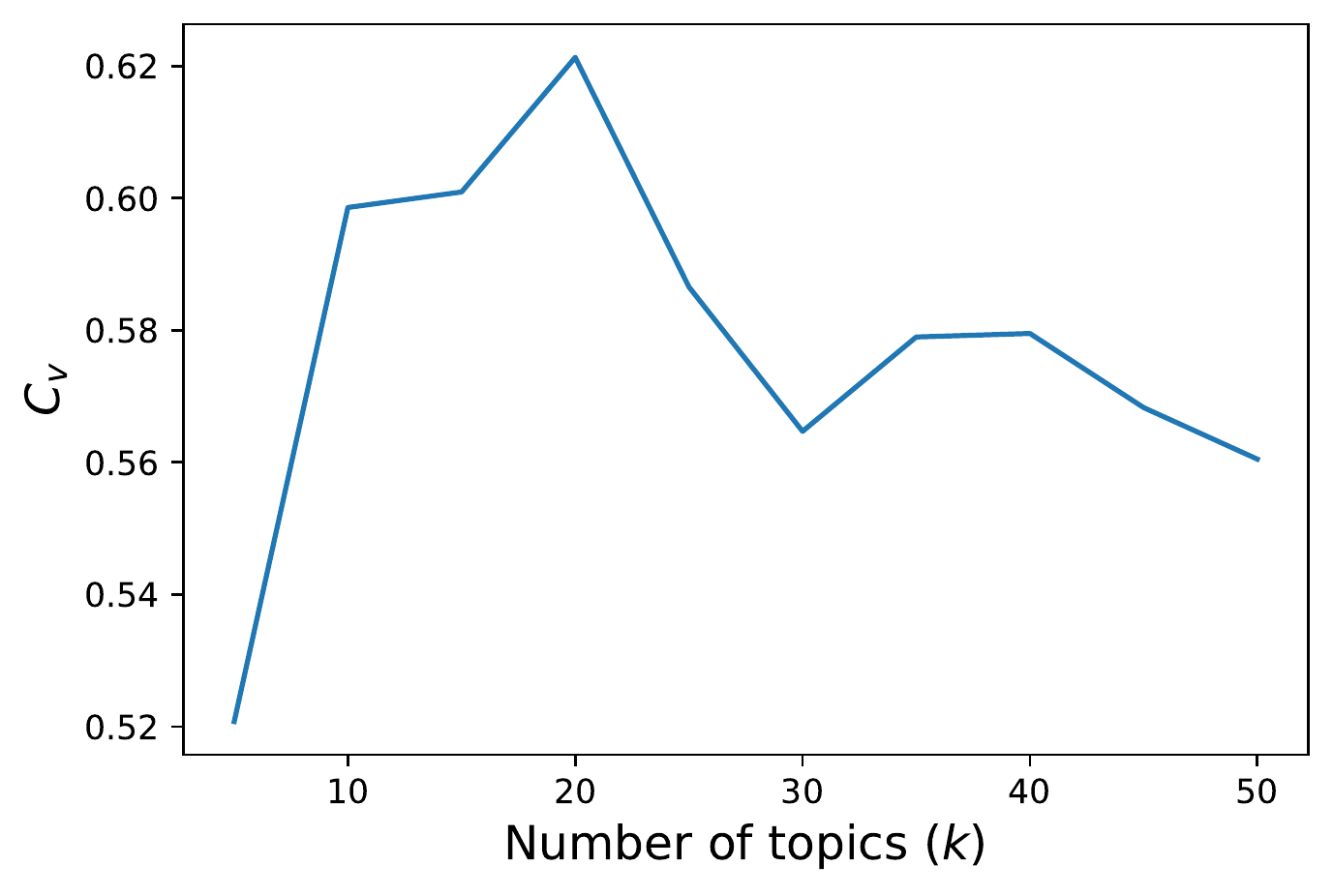}
 \caption{$C_v$ metric according to the number of topics.}
 \label{fig:cv_score}
\end{figure}

In our setting, we consider as a document the aggregate titles of the offered products in each of the 2906 shops in our dataset.
For training LDA models on the generated documents, we employed the implementation provided by Machine Learning for Language Toolkit (MALLET) \footnote{\url{http://mallet.cs.umass.edu/}}.
To obtain the most coherent topic model for our data, we considered the number of topics $(k)$ within the range from 5 to 50 with a step of 5 and trained the LDA models with 1,000 Gibbs sampling iterations and priors $\alpha = 5/k$, $\beta = 0.01$.
For each trained model, we compute the $C_v(k)$ metric 
. This metric combines the indirect cosine measure with the normalised pointwise mutual information (PMI) and the boolean sliding window technique, to determine the number of optimal topic classes according to data distribution \cite{2684822}. According to Figure \ref{fig:cv_score}, the value yielding the highest $C_v$ corresponds to $C_v(20)=0.621$ and thus, we set the number of topics $k$ to 20.

In Table \ref{tab:topics}, we present the topics learned by our best LDA model, including the most relevant terms describing each topic and the number of shops where each topic is dominant. To obtain the most descriptive terms for topic interpretation, we adopted the approach of ranking individual terms within topics presented in \cite{sievert2014ldavis}.

\begin{table}[!th]
 \centering
 \footnotesize
 \def\x{\textbf{Key Terms}}
 \setlength{\tabcolsep}{2pt}
 \rowcolors{2}{gray!25}{white}
\begin{tabular}{cp{3.6in}c}
\toprule
 \textbf{Topic} & \multicolumn{1}{c}{\x} & \textbf{Documents} \\
\midrule
 4 & skin, fortnite, trooper, black, knight, renegade, galaxy, raider, ghoul, recon, expert, ikonik, skull, psn, linkable, aerial, random, assault, season, stack & 466 \\
 2 & spotify, premium, netflix, nordvpn, vpn, family, hulu, pornhub, upgrade, nord, crunchyroll, owner, expire, bulk, uhd, disney, grammarly, screen, student, support & 344 \\
 9 & discord, month, paypal, service, btc, day, buy, week, server, term, read, bot, youtube, twitch, contact, apple, cheap, token, rust, music & 219 \\
 14 & method, amazon, free, balance, paypal, store, acc, guide, money, carding, make, google, ebook, check, work, ebay, refund, script, bank, link & 205 \\
 8 & access, full, minecraft, nfa, minecon, cape, sfa, hypixel, acces, optifine, semi, rank, roblox, unmigrated, vip, mvp, ufa, letter, elite, ninja & 197 \\
 10 & key, lifetime, steam, origin, game, uplay, window, license, edition, office, pro, battlefield, standard, global, year, crypto, fifa, deluxe, microsoft, home & 178 \\
 5 & config, capture, openbullet, checker, fast, cpm, full, ultra, package, proxy, api, onetap, instagram, aimware, master, mycanal, view, cfg, update, real & 153 \\
 11 & premium, hulu, disney, hbo, live, monthly, pass, plan, yearly, commercial, showtime, ad, nba, gold, annual, starz, tidal, fitbit, directv, espn & 135 \\
 13 & private, day, combo, pack, email, mail, pass, combolist, usa, fresh, access, valid, hit, list, user, domain, mixed, shopping, guarantee, database & 130 \\
 3 & point, dominos, balance, wing, wild, buffalo, free, reward, pizza, usa, subscription, jersey, sonic, drive, mike, payment, amc, grubhub, pts, shein & 128 \\
 6 & account, random, gta, shop, dork, crack, site, move, box, source, red, shoppy, mixed, high, target, game, million, keyword, turbo & 119 \\
 1 & premium, year, subscription, vpn, auto, lifetime, renewal, adult, security, site, brazzers, monthly, device, avast, pro, membership, pornportal, unlimited, kaspersky, renew & 96 \\
 16 & method, follower, include, depop, credit, balance, subway, point, free, deliveroo, footasylum, pizza, voucher, refundable, wowcher, tesco, studio, attach, guide, disney & 77 \\
 17 & good, order, android, web, ios, pack, iptv, basic, complete, learn, facebook, website, deezer, test, video, theme, virtual, country, datum, tool & 77 \\
 12 & account, random, level, skin, champion, inactive, euw, legend, champs, verify, league, valorant, unverified, lvl, eune, active, unverfied, lol, region, verfie & 73 \\
 7 & warranty, month, sport, sling, year, orange, blue, pro, dazn, extra, package, viaplay, hbo, pass, usa, nfl, directv, total, disney, xfinity & 68 \\
 15 & account, code, reward, pin, coupon, hellofresh, stock, hotel, pizza, visa, starbucks, ulta, discount, invite, online, american, include, usd, optimum, canada & 65 \\
 20 & card, gift, giftcard, grill, restaurant, pizza, bar, juice, jamba, kitchen, pin, city, read, noodle, fatz, deli, bbq, cafe, event, company & 64 \\
 19 & account, porn, paypal, bitcoin, random, pcoptimum, instagram, twitter, verife, playstation, unverifed, uber, ride, automatic, cyclegear, apple, follower, perfect, lifeselector, likes & 55 \\
 18 & serial, number, gaming, logitech, mouse, wireless, steelseries, pro, monitor, lenovo, inch, model, keyboard, series, gen, notebook, dell, headset, prime, imei & 54 \\
\bottomrule
\end{tabular}
\caption{Different topic classes and their corresponding key terms, sorted according to the number of documents found.}
\label{tab:topics}
\end{table}

To provide an insight on the products sold by the shops classified in each topic, Table \ref{tab:sample_products} includes some indicative examples per topic, with respect to the number of topic-relevant terms contained in their titles.

The latter further allows us to characterise each one of the learned topics in a qualitative manner. Topics \#1 and \#2 describe ``premium'' accounts for a variety of online services and software products including streaming and VPN services.
Topic \#3 describes accounts associated with popular restaurants and fast food companies.
Topic \#4 reflects accounts associated with in-game items and collectables for the popular online game Fortnite. 
This topic is found to be dominant in most shops, in comparison to the other topics, with 466 occurrences (i.e. 16\% of all shops). Although selling game accounts can be perceived as an innocuous activity, provided the context of our data collection, these selling activities could be linked with money laundering schemes, based on the idea of converting stolen money to virtual currencies which are used to purchase in-game items \cite{cloward2020game,moiseienkogaming}.
Topic \#5 focuses on OpenBullet configurations. OpenBullet is a brute-forcing tool used for performing credential stuffing attacks against online services \cite{kirkbride2020game}, which are described by configuration files \textit{``configs''}, offering features such as checking multiple credentials simultaneously (advertised by metrics such as CPM, standing for ``Checks Per Minute'') and bypassing rate-limiting.
Topic \#6 contains several classes associated with a broad spectrum of products ranging from game accounts to hacking and reconnaissance tools such as dorks.
Topic \#7 includes mainly subscriptions to various sports and video streaming services.
Topic \#8 highlights accounts, hacking tools and in-game items for the popular video game Minecraft.
Topic \#9 models the false products previously described (cf Table \ref{tab:false_prods}), containing information regarding vendor's terms of service and links to external Discord servers, Telegram channels, and etc.
Topic \#10 includes product licences and keys for a variety of software packages, games and operating systems.
Topic \#11 describes subscription plans for streaming services, similar to Topic \#7.
Topic \#12 involves accounts for the popular game League of Legends.
Topic \#13 describes selling leaked user data from security breaches, in the form of \textit{combo lists}, i.e. combinations of usernames/emails and passwords \cite{malderle2018gathering}, which can be used for compromising accounts with the same credentials in other services, by means of credential stuffing attacks, as seen in Topic \#5.
Topic \#14 involves mostly guides and e-books regarding carding and other methods of financial fraud.
Topic \#15 contains discount codes and accounts containing redeemable credits for various online shops and e-commerce platforms.
Topic \#16 is closely related to topics \#14 and \#15 and includes vouchers for online purchases in various venues as well as methods to perform a fraud or to scam sellers.
Topic \#17 mainly includes subscriptions for online services and products with a focus on mobile apps.
Topic \#18 is related to serial numbers for computer peripherals such as monitors, keyboards, etc.
Topic \#19 provides assorted \textit{``random''} accounts for various social media and sites.
Finally, Topic \#20 is related to products such as redeemable gift cards, mainly for restaurants and food suppliers.

\subsection{Use case analysis of surface data}
\begin{table}[!th]
\setlength{\tabcolsep}{2pt}
 \centering
 \tiny
  \def\y{\textbf{Sample Products}}
  \renewcommand*{\arraystretch}{1.1}
\resizebox{\textwidth}{!}{
\begin{tabular}{cp{2in}|cp{2in}}
\toprule
 \textbf{Topic} &\multicolumn{1}{c}{\y}  & \textbf{Topic} & \multicolumn{1}{c}{\y} \\
\midrule
1 & Avast Premier Premium Security 2 Year 1 Device & 11 & Hulu | Premium | Plan - No Commercials, ShowTime, Live TV, HBO, Cinemax, STARZ, Entertainment Add-on. \\
1 & HMA VPN | PREMIUM | AUTO RENEWAL | MONTHLY/YEARLY | KEY/ACCOUNT & 11 & NBA League Pass Premium | Monthly Subscription | 1 Week Warranty \\
1 & Spotify Premium Account + [Auto-renewal] & 11 & Disney+ | Bundle Monthly Plan Hulu, ESPN+ \\

2 & Nord Vpn Premium | Expire - 2021 & 12 & EUNE Verfied Inactive (IRON) Level 30 Account Random Champs \& Skins \\
2 & Netflix Premium Accounts (UHD) & 12 & EUW Level 30+ Verfied [Silver Kayle] Account Random Champs \& Skins \\
2 & Spotify Family Owner Premium & 12 & NA Mystery Account Level 30 Inactive verfied RANDOM Everything 0-Max Champs 0-900 Skins \\

3 & Jersey Mike's Free Regular Sub - Wrap - Tub USA & 13 & 151k USA Valid Mail Access Combolist HQ Private \\
3 & Dominos 2 Free Pizza (USA) & 13 & 1.1 Million USA Domain Valid Mail Access Combolist Private \\
3 & Buffalo Wild Wings (USA) (1500-2000pts) & 13 & 110GB BRAZZERS USA DATABASE HQ [USER:PASS] COMBO \\

4 & SPECIAL OFFER RENEGADE RAIDER+160SKINS & 14 & [EBOOK] Amazon Carding Giftcards Pro Method 100\% Work \\
4 & Fortnite I Renegade Raider + Recon Expert + Black Knight & 14 & Paypal: Double Your Balance [Method][Guide] \\
4 & Fortnite account with 3 EPIC SKIN skull trooper + ikonik + the ace Warranty 100/100 & 14 & Free .RDP for Paypal Carding Method \\

5 & CONFIG INTERMARCHE FR API FULL CAPTURE ULTRA FAST CPM (socks4/5) & 15 & HelloFresh \$20 Discount Code (PayPal) \\
5 & Subway Config + Full Capture for OpenBullet (Fast CPM) & 15 & DISCOUNT CODE 20\% - 25\% ADIDAS US \\
5 & [OpenBullet] STREAMATE API CONFIG [ FULL CAPTURE ] & 15 & Starbucks. ACCOUNT with \$6.50, 200 stars \\

6 & == Depop Account 10k Followers == [HQ SHOP] & 16 & Deliveroo Refundable- £30.00- £34.99 -48HR (Method Included) \\
6 & GTA V Account (Cr4ck3d) & 16 & Deliveroo Free Food Method (In Depth) \\
6 & 10x Hulu Account Random Subscription & 16 & FootAsylum - Account Balance - £8+ \\

7 & Sling Orange \& Blue + Sports Extra + NBA League Pass | 6 Months Warranty & 17 & Scribd | Read Books, Audiobooks \& Magazines - 1 year warranty [Web/iOS/Android] \\
7 & DAZN USA | 1 Year Warranty & 17 & SkillShare.com Premium - 3 months warranty [Android/iOS/Web] \\
7 & Hulu Premium | 1 year warranty (Package: No Commercials) & 17 & Deezer [Android/iOS/Web] - 1 year warranty \\

8 & Minecraft unmigrated full access account - With optifine cape & 18 & Logitech PRO Wireless Gaming Mouse (Serial Number) \\
8 & DMC 2.1 - Minecraft Checker / VIP HYPIXEL, CAPE OPTIFINE, CAPE MINECON, SECURED , INSECURED & 18 & HP EliteDisplay E223 21.5-inch Monitor Serial Number \\
8 & Hypixel VIP+ Account [Lifetime] Minecraft Non-Full-Access & 18 & SteelSeries Arctis Pro Wireless Serial \\

9 & Minecraft FA Account (To buy with Paypal contact us on Discord!) & 19 & PlayStation Account 5-10 Random Games \\
9 & Discord Token Checker [BOT] FREE READ DESC & 19 & Instagram Random Account \\
9 & Contact me / discord server & 19 & Twitter Random Account x5 \\

10 & Borderlands 3 Standard Edition Epic Games Key & 20 & Chipotle Gift Card \$10-\$20 [Pin less] \\
10 & Microsoft Office 2019 Pro Plus (1 PC License) & 20 & Round Table Pizza \$40 Gift card + PIN \\
10 & Windows 8 PRO Digital License Key 32 \& 64 Bit & 20 & Farrelli's Pizza Gift Card 50\$ Giftcard \\
\bottomrule
\end{tabular}
}
\caption{Sample products for each topic.}
\label{tab:sample_products}
\end{table}

In this section, we focus on the Topics \#5 and \#13, which as highlighted above, model products related with cybercriminal activities such as selling breached credential dumps and using tools for automating the compromise of accounts in different online services.
To this end, we leverage the term-salience metric defined in \cite{chuang2012termite}, which given the set of representative terms per topic, ranks them according to their distinctiveness, i.e. how informative a specific term is for determining the generating topic, versus a randomly-selected term. Subsequently, we select the top-3 most salient terms for topics \#5 (\textit{config}, \textit{openbullet}, \textit{capture}) and \#13 (\textit{combo}, \textit{database}, \textit{records}), and we use them to query product titles, in order to identify the most prevalent products modelled by these topics. For Topic \#13, we additionally include the term \textit{db} which is a common abbreviation for the term \textit{database}.  

As previously reported (Table \ref{tab:sample_products}), Topic \#13 models leaked data from online data breaches, which are sold in the form of username/email and password combinations, along with other personal information. Such listings usually advertise the number of the breached records, as well as the source of the leak. In Table \ref{tab:breaches} we present some of the largest account dumps found in our \texttt{Shoppy} dataset, along with their prices. Indeed, we discovered that popular password breaches checker platforms, such as \url{https://haveibeenpwned.com}, list the majority of the account database dumps sold on \texttt{Shoppy}. Moreover, this could explain the relatively low price tag for leaks, including up to millions of records, as the respective breaches have already been made public.

In Table \ref{tab:openbullet}, we list some illustrative products with titles including at least one of the selected salient terms for Topic \#5. We observe that these products represent configurations for software such as OpenBullet\footnote{\url{https://github.com/openbullet/openbullet}}/BlackBullet\footnote{\url{https://redskyalliance.org/xindustry/blackbullet-credential-stuffing}}/Storm\footnote{\url{https://www.netacea.com/blog/storm-cracker-tool/}}. As previously stated, such tools can be used to automate credential stuffing attacks \cite{rees2020credential}, versus various online services, as shown from the product titles. Sellers of such ``configs'' often advertise features such as CPM (checks per minute) and capturing functionality offered, i.e. the ability to capture specific information associated with a compromised account, such as saved credit cards and payment methods, reward points, etc.

From the above, we can largely infer the modus operandi of the account sellers of \texttt{Shoppy} and other cybercriminal markets: One is able to purchase massive quantities of breached credentials, and by exploiting the password reuse behaviour exhibited by many users \cite{poornachandran2016password}, she could compromise users accounts with same credentials in other online services by using credential stuffing tools with different configurations.


\begin{table}[!th]
 \centering
 \scriptsize
 \rowcolors{2}{gray!25}{white}
\begin{tabular}{p{3.5in}rrr}
\toprule
 \textbf{Title} & \textbf{\# Records} & \textbf{Price} \\
\midrule
 Combo List | 528M Yahoo.com & 528,000,000 & 400 \\
 Combo List 376M Hotmail.com & 376,000,000 & 200 \\
 Facebook - 267 Million Records Breach [FULL DB] & 267,000,000 & 500 \\
 Combo List 258M Gmail.com & 258,000,000 & 250 \\
 Zynga - 213 Million Records & 213,000,000 & 250 \\
 Dubsmash Full Database | 162 million | hashed & 162,000,000 & 120 \\
 DubSmash - 162 Million Records (FULL SQL DB) & 162,000,000 & 35 \\
 MyFitness Pal - 4.62GB (144 Million Records) & 144,000,000 & 25 \\
 Xiaomi - 144 Million Records & 144,000,000 & 25 \\
 Canva - 137 Million Records & 137,000,000 & 80 \\
 111 MILLION-RECORD PEMIBLANC USA DATABASE COMBO LIST & 111,000,000 & 30 \\
 MyHeritage - 92.2 Million Records & 92,200,000 & 75 \\
 Houzz - 49 Million Records & 49,000,000 & 300 \\
 Facebook DB - 45 Million & 45,000,000 & 265 \\
 Chegg - 29 Million Records (Dehashed) & 29,000,000 & 50 \\
 Evony - Multiplayer Game : 28.7 Million + 13.8 Million Records & 28,700,000 & 24 \\
 Hautelook - 28 Million Records (Full DB) & 28,000,000 & 100 \\
 24 Million LUMINATI PROXY DATABASE HQ [EmailPass] Combo & 24,000,000 & 40 \\
 YouNow - 18.2 Million Records & 18,200,000 & 50 \\
 8tracks - 18 Million Records & 18,000,000 & 41 \\
 500PX - Full DB [14.9 million records] & 14,900,000 & 250 \\
 Dubsmash | 12 million lines | Private Combo & 12,000,000 & 25 \\
 CouponMom / Armor Games - 11 Million Records & 11,000,000 & 41 \\
 Cafepress *NEW* 11 Million Records & 11,000,000 & 41 \\
 Bitly - 9.3 Million Records & 9,300,000 & 41 \\
 BlankMediaGames - 7.6 Million Records Breach & 7,600,000 & 41 \\
 GAMESTOP.COM Database | 7.6M UHQ LINES | 100K Splits | Mail:Pass | Good for EVERYTHING & 7,600,000 & 10 \\
 StockX - Full 6.8 Million Records DB & 6,800,000 & 65 \\
 SNAPCHAT.COM | DATABASE | LEAK | 4,6M LINES | PHONE NUMBERS, USERNAMES | & 6,000,000 & 2 \\
 5.7M Facebook Profiles w/Email & 5,700,000 & 20 \\
 Stronghold Kingdoms - 5.1 Million Records & 5,100,000 & 50 \\
 5M BITLY DATABASE HQ COMBOLIST (Netflix,Hulu,Spotify,PSN,\& More) & 5,000,000 & 20 \\
 4.6 Million Snapchat.com Databases Private Combos & 4,600,000 & 20 \\
 Game Salad [Dehashed] - 1.8 Million Records & 1,800,000 & 65 \\
 PRIVAT HQ 800k Yahoo.com USA Combolist & 800,000 & 20 \\
 Hookers.nl (Dutch prostitution forum) - 291K Records & 291,000 & 50 \\
 240k Sbcglobal.net Domain HQ private Combolist 100\% & 240,000 & 8 \\
 225k Icloud.com Domain HQ Private Combolist 100\% HQ & 225,000 & 12 \\
 211k Cox.net Domain HQ Combolist 100\% Private Base & 211,000 & 12 \\
 Coinmama dehashed db 209k mail:pass. exclusive & 209,000 & 700 \\
 naughtyamerica.com [PORN] {Mail:Pass} 114K Database & 114,000 & 0 \\
 110k Sbcglobal.net Domain HQ private Combolist 100\% & 110,000 & 8 \\
 100K ONLY YAHOO COMBOLIST - EMAIL-PASS & 100,000 & 3 \\
 100K HQ ONLY HOTMAIL COMBOLIST - EMAIL-PASS & 100,000 & 3 \\
 51k Cox.net Valid Mail Access Private Combolist & 51,000 & 12 \\
 39k Cox.net Domain HQ Private Combolist & 39,000 & 7 \\
\bottomrule
\end{tabular}
\caption{Known breaches sold through Shoppy, as identified by Topic \#13's most salient terms.}
\label{tab:breaches}
\end{table}

\begin{table}[!th]
 \centering
 \footnotesize
 \rowcolors{2}{gray!25}{white}
\begin{tabular}{p{4in}r}
\toprule
\textbf{Title} & \textbf{Price} \\
\midrule
Custom Config for Openbullet & 999 \\
New Nectar Card capture Site \#2 config for Blackbullet & 450 \\
nintendo captchaless open bullet config & 250 \\
Spotify | Config [OpenBullet] & 200 \\
Skout Dating Site OpenBullet config [Anom] & 100 \\
Badoo.com OpenBullet config (Fixed on 30.1.2020 & 100 \\
Benaughty \& Naughtydate Openbullet Configs(2 configs) & 100 \\
NINTENDO SWITCH CONFIG [WITH FULL GAME CAPTURE PAYMENT METHOD AND BALANCE] & 100 \\
Luminati.io OB config & 50 \\
{[OPENBULLET]} COOP UK CONFIG WITH FULL CAPTURE & 50 \\
CONFIG MYCANAL API FULL CAPTURE + CHECK MAIL ACCESS ULTRA FAST CPM & 40 \\
PSN Captchaless API (50K CPM) Config | Full Capture & 40 \\
Btc.com BlackBullet Config with CAPTURE & 35 \\
Custom Config ( We code your Configs, Web applications, Scrapers, Bruteforcers, Everything related ) & 	20 \\
{[OPENBULLET]} KrispyKreme Config | With Detalied Captures & 20 \\
{[OPENBULLET]} Grubhub Config | CAPTURES CC, PAYPAL, AND GC BALANCE & 15 \\
NordVPN Config + Expiration Capture & 15 \\
{[OPENBULLET]} Papa John's Config | CAPTURES POINTS & 15 \\
CodeCademy API Checker | +2.6k CPM | Capture: Pro & 15 \\
Apple Valid Emails Checker By OPENBULLET GROUP & 15 \\
{[CCShop]} streetcc.pw *.loli for OpenBullet [Capture Balance] & 15 \\
Facebook Config Capture (check if its ad account or not) Very Fast & 15 \\
{[OB]} Config ICams.com With Capture Balance & 10 \\
{[OPENBULLET]} McDonalds USA Config | CAPTURES CC & 10 \\
SkinHub | +8k CPM | Captures: [Balance,RefBalance,Country,TotalWithdrawals,TotalDeposits] & 10 \\
{[OPENBULLET]} Wiki Mining BTC Config + Capture & 5 \\
{[OB]} PicArt With Capture Followers & 5 \\
{[openbullet]} Shipt - [High CPM] Orders + Cards + Rewards Program Capture & 5 \\
{[STORM]} CONFIG NETFLIX + FULL CAPTURE [WORKING] \& FAST & 3 \\
{[STORM]} Hotstar Config + Capture (Fast) & 2 \\
\bottomrule
\end{tabular}
\caption{Indicative products modelled by Topic \#5 in descending price order.}
\label{tab:openbullet}
\end{table}

\section{Discussion and final remarks}

There are several conclusions that can be extracted from the analysis and the outcomes obtained in the previous sections. First and foremost, we found evidence of malicious activities which are usually taking place in the dark web, yet this time arising on the surface web. In this sense, the cynicism of malicious actors, who are perpetrating these activities, is covered by a lack of methodologies and takedown mechanisms, due to several factors such as, e.g. the decentralised nature of the marketplaces. To the best of our knowledge, this is the first work that provides a solid and automated methodology to find, quantify and classify in a comprehensive way such activities. Nevertheless, despite the promising analysis leveraged in this article, malicious actors always find a way to circumvent analysis, since, e.g. they only use such platforms as a contact point, redirecting all of their activities to other external channels such as Telegram or Discord. Moreover, the use of technologies such as IPFS can augment the possibilities and resilience of such malicious practices \cite{ijisipfs,POLITOU2020956}.

Another dimension to be yet explored is the underlying connection between the activities reported in this article and further criminal campaigns. Therefore, despite the fact that most of the sold products can be classified as `soft' cybercrime (i.e. passwords, credit card credentials, personal data) they can pose significant damage to individuals and businesses, and they may be just the tip of the iceberg. More concretely, money laundering and the financing of other, probably more dangerous activities, can be just happening in front of our eyes \cite{irwin2010detecting,mikhaylov2016cards,dacosta2020cybercrime}.

As previously stated, there exist several challenges for the analysis and takedown of the illegal activities being hosted on such platforms. The decentralised nature of, e.g. \texttt{Shoppy} avoids crawling mechanisms that could be used to collect all the stored information. Moreover, \texttt{Shoppy} is a resilient platform, as well as \texttt{Sellix} and \texttt{Selly}. The latter is supported by the fact that users can just have back their shops easily. As a matter of fact, the activities reported in this article are taking place at the moment of writing without restrictions.

The possibility of linking these activities by using novel blockchain platforms is a further issue that needs to be thoroughly explored. First, the immutable nature of blockchain may permit the development of shopping platforms which offer private and permanent selling services \cite{8666470,LIU2020102059,8879578}. The latter fact is a critical issue due to the lack of efficient erasure mechanisms \cite{ateniese2017redactable,politoublock,8941045}.

We argue that more effort should be devoted to the development of robust AI methods as well as data collection procedures such as the one proposed in this article to locate and quantify the extent of such activities. Moreover, robust investigation protocols and more support from law enforcement towards the prosecution of these activities, as well as legislation related to this phenomena are mandatory. Finally, proactive measures, including strategies such as abnormal behaviour detection and the corresponding mitigation actions should be implemented by design, especially in the cases in which a platform is using immutable architectures.

In this article, we showed that most of the activities that are leveraged in the dark web are also taking place on the surface web and yet, no effective mechanisms or takedown measures are taking place. This claim is supported by our thorough analysis of a marketplace, namely \texttt{Shoppy}. First, we collected credentials from two well-known forums, namely cracked and blackhatworld. Next, due to decentralised and anonymous nature of \texttt{Shoppy}, we used such credentials to crawl and retrieve data regarding shops, products and descriptions. Subsequently, we used topic modelling-based analysis to categorise and further explore the collected data by reporting several qualitative and quantitative features. Our findings evince the cybercriminal nature of a myriad of shops and users in the \texttt{Shoppy} ecosystem, supporting our initial claim. Finally, to raise awareness and highlight the relevance of our findings, we discussed the implications of our research, the current challenges and limitations, and proposed some measures to overcome them. Future work will focus on exploring similar marketplaces and trying to find correlations between different platforms in an automated way. Moreover, we plan to analyse the possible links between the activities leveraged in such marketplaces and cryptocurrencies, as well as other widely used financial platforms.

\section*{Acknowledgements}
This work was supported by the European Commission under the Horizon 2020 Programme (H2020), as part of the projects \textit{CyberSec4Europe} (Grant Agreement no. 830929) and \textit{LOCARD} (Grant Agreement no. 832735).

The content of this article does not reflect the official opinion of the European Union. Responsibility for the information and views expressed therein lies entirely with the authors.

\bibliographystyle{plain}
\bibliography{references}
\end{document}